# Giant Thermal Rectification from Polyethylene Nanofiber Thermal Diodes

Teng Zhang[†] and Tengfei Luo[*,†,‡]

[†]Aerospace and Mechanical Engineering and [‡]Center for Sustainable Energy at Notre Dame, University of Notre Dame, Notre Dame, Indiana 46556, United States

## ABSTRACT

The realization of phononic computing is held hostage by the lack of high performance thermal devices. Here we show through theoretical analysis and molecular dynamics simulations that unprecedented thermal rectification factors (as large as 1.20) can be achieved utilizing the phase dependent thermal conductivity of polyethylene nanofibers. More importantly, such high thermal rectifications only need very small temperature differences (< 20 $^{\circ}$C) across the device, which is a significant advantage over other thermal diodes which need temperature biases on the order of the operating temperature. Taking this into consideration, we show that the dimensionless temperature-scaled rectification factors of the polymer nanofiber diodes range from 12 to 25 – much larger than other thermal diodes (< 8). The polymer nanofiber thermal diode consists of a crystalline portion whose thermal conductivity is highly phase-sensitive and a cross-linked portion which has a stable phase. Nanoscale size effect can be utilized to tune the phase transition temperature of the crystalline portion, enabling thermal diodes capable of operating at different temperatures. This work will be instrumental to the design of high performance, inexpensive and easily processible thermal devices, based on which thermal circuits can be built to ultimately enable phononic computing.

* Address correspondence to tluo@nd.edu

# 1. Introduction

The unparalleled development of information technologies originates from the inventions of several critical electronic devices which can manipulate electron flow, exemplified by the solid-state electronic diode.[1] While electronics dominate today's information technologies, their counterparts – phononics, which also have the potential to transmit and process signals, have not yet been realized in applications. Instead of controlling electron flow, phononic devices control thermal energy (heat) flow, which is much more abundant than electrical energy. Over 90% of the world energy production is related to heat flow with only around 13% converted into useful electricity and even less used for electronics.[2] Due to the abundance of thermal energy, the realization of phononics may result in even more wealthy applications than electronics.

Thermal rectification is the key for heat flow control.[3-5] A device that rectifies heat flow is called a thermal diode which allows larger heat flux in the forward direction than that in the reverse direction given the same temperature difference. The performance of rectification can be quantified by a factor which relates the forward ($\kappa^+$) and the reverse ($\kappa^-$) thermal conductivities (or heat flux):[5]

$$\varepsilon = \frac{\kappa^+}{\kappa^-} - 1 = \frac{J^+}{J^-} - 1 \qquad (1)$$

where $J^+$ and $J^-$ are the heat fluxes in the forward and the reverse cases, respectively. The rectification factor is defined as a non-negative value, with 0 meaning no rectification effect.

Thermal rectification has been realized in early years using bulk materials. In bulk materials, thermal rectifications are observed exclusively in bi-material junctions, with the first observation made by Starr in 1936 on a metal-nonmetal junction.[6] Most measured rectification factors in bulk junctions are much lower than 0.5, and huge temperature differences (~O(100 K)) are needed to achieve appreciable rectification above 0.5.[5-12] Theoretical predictions have shown that a temperature difference on the same order as the average temperature is needed to achieve rectification factors higher than 2, because thermal conductivities are usually weak functions of temperatures.[13] This limitation largely prevents the practical applications of bulk thermal diodes. Dames[13] proposed a dimensionless temperature-scaled rectification factor which takes into account the impact of the magnitude of temperature biases (Eq. 2). The scaled rectification factors of these bulk diodes are all below 8.

$$\hat{\varepsilon} = \frac{\varepsilon}{2(T_H - T_L)/(T_H + T_L)} \qquad (2)$$

In the past decade, theories and simulations have shown that nanostructures are promising platforms to achieve high performance thermal diode,[14,15] inspiring a renewed interest in thermal rectification.

Nano-structures are favored over bulk materials for many applications since they miniaturize the devices and reduce material cost. Besides applications in phononics, nanoscale thermal rectification can also play important roles in thermal management of microelectronics[16] and energy storage.[17-19] Using atomistic simulations, asymmetric carbon structures such as carbon nanotubes and graphene are shown to have rectification factors ranging from 0.1-3.5.[20-26] However, all the proposed devices with a rectification factor above 0.5 need huge temperature differences (~O(100 K)). Due to this reason, these proposed diodes have rather low $\hat{\varepsilon}$ (<7). Very limited experimental results exist for nanoscale thermal diode. Chang et al.[27] observed rectification in a carbon nanotube coated asymmetrically with molecules, but the achieved rectification factor was rather low (0.07) with a ~20 K temperature difference. By utilizing the interplay between metallic and insulating phases in vanadium oxides, Zhu et al.[28] reported a solid-state temperature-gated thermal rectifier, and the largest achieved rectification factor was 0.28. The mechanism of this diode is not completely clear. To our best knowledge, there is no thermal diode with rectification factors larger than 0.5 with small temperature biases.

Recently, Kobayashi and coworkers[29] utilized bulk phase change materials (PCM) to form a bi-material junction and observed thermal rectification around 55 K with a temperature difference as low as 2 K. Their device makes use of the sharp thermal conductivity change due to the phase transition of one portion of the junction. The achieved rectification factor, however, is very low (0.14) due to the small contrast ratio of the thermal conductivities in the two phases. Recently, atomistic simulations show that a very sharp thermal conductivity change by almost one order of magnitude can be achieved in polyethylene nanofibers (PENF) around 390 K.[30] Such a sharp thermal conductivity change is related to phase transition and is fully reversible. The phase transition temperature can also be tuned by changing the diameter of the fibers.[31] These features make PENF a promising component for high performance thermal diodes.

In this paper, we propose and study a bi-material fiber junction by connecting a crystalline PE fiber to a cross-linked PE (PEX) fiber to realize very high thermal rectification factors. We show through analytical study and molecular dynamics (MD) simulations that such thermal diode designs can achieve rectification factors as high as 1.20 with a temperature bias of only 20K. We also demonstrate that changing the fiber diameter, which changes the phase transition temperature of the crystalline portion of the bi-material junction, enables the design of thermal diodes that can operate at various temperatures.

## 2. Results and Discussion

### 2.1 Design of thermal diode

Though amorphous polymers are usually thermal insulators due to their low thermal conductivity (0.1–1W/mK),[32] recent experiments show that crystalline PE fibers can have thermal conductivity in the range of 20-100 W/mK.[33-37] However, the thermal conductivity of PE fibers is very sensitive to the phase, especially the rotation of the chain segments. Our previous MD studies show that at 397 K a morphology change from an all-*trans* conformation crystalline structure to disordered structure with segmental rotations can happen, which leads to a sharp change in the thermal conductivity (**Figure 1**a), and this change is fully reversible.[31] Such a sharp, high contrast and reversible thermal conductivity switching serves as the basis for creating a high performance thermal diode.

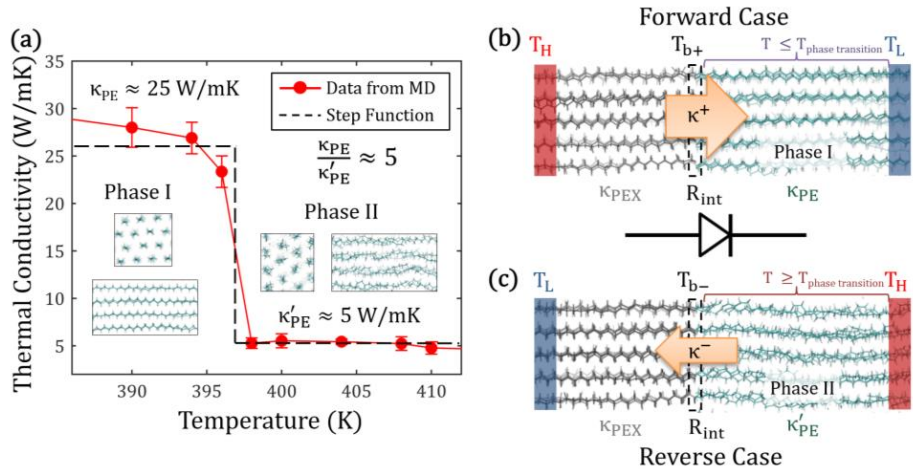

**Figure 1.** a) Sharp thermal conductivity change of PENF due to phase transition. Schematics of PE-PEX thermal diode junction with b) forward and c) reverse temperature biases.

Our proposed thermal diode consists of a crystalline (linear) PE fiber connected to a PEX fiber end-to-end (Figure 1b). Actually, such a bi-material junction can be easily fabricated. The simplest practical strategy is to convert a portion of the polymer fibers into cross-linked polymer. Such conversion can be realized by exposing part of the polymer fiber to a cross-linking agent like vinylsilane, an ultraviolet light or an electron beam irradiation.[38,39] The linear PE fiber portion has the above-mentioned thermal conductivity switching effect due to phase transition at a specific temperature, while the PEX part has thermal conductivity that is not sensitive to temperature since the cross-linking largely restrains atomic movements.

## 2.2 Theoretical thermal rectification

Figures 1b & 1c show two schematics of the thermal diode subject to opposite temperature biases. When the linear end is subject to low temperature (Figure 1b), it is in the high thermal conductivity phase, and the effective thermal conductivity of the junction is high. When the temperature bias is reversed (Figure 1c), the high temperature at the linear end enables phase transition which lowers the thermal conductivity, leading to a lower effective thermal conductivity of the junction. As long as the operating temperature and the length of each portion in the junction are carefully designed to ensure that the linear portion undergoes phase transition when the temperature bias is reversed, thermal rectification can be achieved.

To find the proper PE/PEX length ratio to achieve the optimal rectification factor, theoretical analysis is first performed based on the thermal conductivity of the linear PE shown in Figure 1a (see Section S1 in Supporting Information (SI) for detailed analysis). One major advantage of utilizing the strong temperature-dependent thermal conductivity in PE is that we can operate the thermal diode with a small temperature bias. In this work, we use a temperature bias of 20 K. Within such a small temperature range, the thermal conductivity within one phase is approximated as a constant (25 W/mK for phase I and 5 W/mK for phase II) and the sharp change is approximated as a step function (Figure 1a).

It can be shown that the thermal rectification factor in Eq. (1) can be expressed in the following form:

$$\varepsilon = \frac{\gamma - 1}{\alpha + 1} \qquad (3)$$

where $\alpha = \frac{R_{int} + R_{PEX}}{R_{PE}}$ is defined as the thermal resistance ratio with $R_{PE} = \frac{L_{PE}}{\kappa_{PE}}$, $R_{PEX} = \frac{L_{PEX}}{\kappa_{PEX}}$ and $R_{int}$ being the thermal resistance of the PE portion, the PEX portion and the interface, respectively. $L_{PE}$, $L_{PEX}$, $\kappa_{PE}$ and $\kappa_{PEX}$ are the lengths and thermal conductivity of the linear PE and PEX, respectively. $\gamma = \kappa_{PE}/\kappa'_{PE}$ is the contrast ratio of the thermal conductivities of the linear PE before ($\kappa_{PE}$) and after ($\kappa'_{PE}$) phase transition and $\kappa_{PE} > \kappa'_{PE}$. It is apparent that a larger $\gamma$ will result in a larger rectification factor, $\varepsilon$. For a given $\gamma$, $\varepsilon$ will increase as the resistance ratio, $\alpha$, becomes smaller (**Figure 2**a, red line). As a result, to achieve large $\varepsilon$, $\alpha$ should be minimized. However, $\alpha$ cannot be infinitesimal. To achieve rectification, the linear PE portion must be in different phases when the heat flux direction is flipped. Thus, the temperature at interface in the forward-case ($T_{b+}$) needs to be below the phase transition temperature to maintain the crystalline phase in the entire linear PE portion, and that in the reverse case ($T_{b-}$) needs to be above the phase transition temperature for the disordered PE phase to occur. As a result, these conditions impose the constraint of $T_{b-} \geq T_{phase\ transition} \geq T_{b+}$ and can be written as Eq. (4). This sets a lower limit to $\alpha$.

$$\beta = \frac{T_{b-} - T_{b+}}{\Delta T} = \left(\frac{\alpha}{\alpha+\gamma} - \frac{1}{\alpha+1}\right) \geq 0 \qquad (4)$$

where $\Delta T = T_H - T_L$ is the temperature bias applied to the junction. As shown in Figure 2a, it is obvious that the rectification factor is maximum ($\varepsilon_{max}$) when $\beta \to 0$, i.e., $\alpha$ reaches its smallest allowable value, $\alpha_{min}$. If the interfacial thermal resistance ($R_{int}$) is ignored, the length ratio between the linear PE ($L_{PE}$) and the PEX ($L_{PEX}$) portions to achieve the maximal thermal rectification can be calculated:

$$\left(\frac{L_{PE}}{L_{PEX}}\right)_{optimized} = \frac{1}{\alpha_{min}} \times \frac{\kappa_{PE}}{\kappa_{PEX}} \qquad (5)$$

It is reasonable to ignore $R_{int}$ since the PE-PEX interface is connected by strong covalent bonds which are expected to present very small thermal resistance. This is also justified in our MD simulations in the following context.

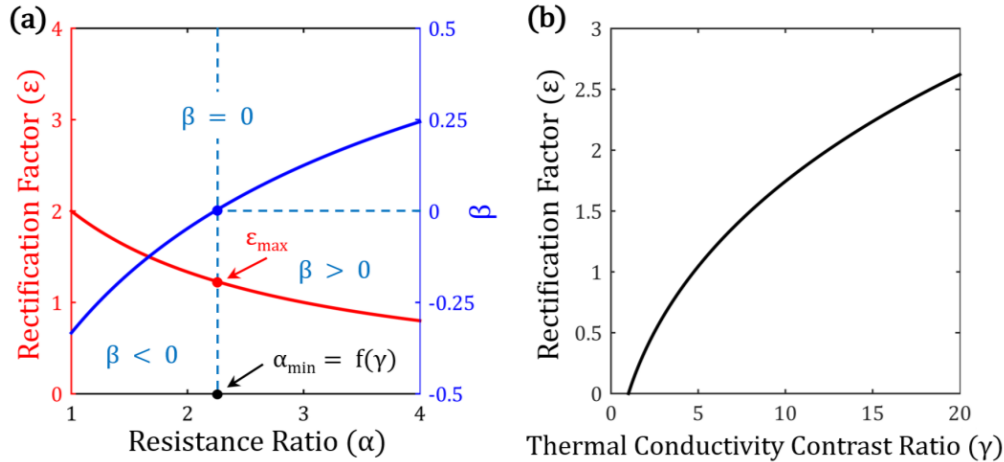

**Figure 2.** Predicted rectification factor, $\varepsilon$, as a function of a) resistance ratio, $\alpha$, with different conditions in $\beta$, and b) the contrast ratio, $\gamma$, of the thermal conductivity change in a phase transition.

Theoretically, $T_{b-} - T_{b+}$ can be infinitesimal since phase transition is discontinuous in temperature. However, practically, $T_{b-} - T_{b+}$ needs to be finite. As mentioned before, for practical applications, we also require the temperature bias, $\Delta T$, to be small, which has been a barrier for the application of previously proposed thermal diodes.[13] If we assume $T_{b-} - T_{b+} = 2K$, which is a large enough window for phase transition,[40] and $\Delta T = 20K$, which is a reasonably small temperature bias, in other words, $\beta = 0.1$, we can obtain a rectification factor of 1.04, given the ~5 times thermal conductivity switching observed in PENF in Figure 1a.[31]

For the given $\beta = 0.1$, the rectification factor as a function of the thermal conductivity switching ratio, $\gamma$, is plotted in Figure 2b. It is worth noting that our previous study has reported that the thermal conductivity switching factor predicted from MD was affected by the finite simulation size.[31] If such finite size effect was removed, the switching factors can be much larger than ~5. Moreover, the above analysis has not taken into account the gradual thermal conductivity change away from the phase transition temperature and the possible interfacial thermal rectification effects. If these factors are all considered, the thermal rectification factor will be even larger than that calculated here. However, even the above predicted rectification factor of 1.04 is 8-fold larger than that achieved using bulk PCM,[29] 6.5 times larger than that from the carbon nanotube-based thermal diode,[25] and 3.7 times larger than the recently demonstrated $VO_2$-based thermal diode.[28] In the next section, we use MD simulations to include the above mentioned factors and mimic the realistic situations to validate the predicted thermal rectification in PE-PEX junctions.

## 2.3 Molecular dynamics simulations of thermal rectification

### 2.3A Thermal rectification in optimized design

In the MD simulations, we study a bi-material fiber junction closely mimicking the proposed PE-PEX junctions. The simulation conditions are set to simulate realistic situations as closely as possible. Non-equilibrium MD (NEMD) is used to calculate the thermal properties. The standard details of the simulations can be found in the Method Section. In these studies, fibers with finite diameters are stimulated by leaving enough vacuum surrounding the fiber. The fiber diameters studied are 2.83 nm, 3.74 nm, 5.29 nm and 7.48 nm. The total lengths of the fibers are around 50 nm (400 -$CH_2$- segments), and those of each portion are determined according to Eq. (5). However, the finite length, 50 nm, exaggerates the importance of the PE-PEX interface: due to the irregularity and larger cross-section in the structure of the PEX portion at the interface, the order of the PE portion near the interface, which is connected to the PEX part chain by chain, will be affected. Since the PE portion is relatively short, such an effect is found to influence a significant portion of the linear PE, making the well-aligned crystal structure hard to maintain. However, this problem should not exist in realistic fiber junctions which can be much longer to provide a transition region from disordered structure to well-ordered PE. Our computational resources, however, prevent us to perform all the simulations on fibers much longer than 50 nm. Thus, it is important to find a way to model the PEX portion with lattice similar to that of linear PE, so that the crystalline structure of linear PE portion is not disturbed by such artifacts.

While understanding that the use of PEX is due to the fact that it is practically easy to fabricate and that its functionality is to prevent phase transition, we can actually use any material that does not undergo phase transition in the operating temperature range of the thermal diode. As a result, we mimic the "cross-linked" PE portion by applying artificial constraints on the carbon atoms of the pristine PE fiber to prevent significant thermal expansion and thus suppress phase transition. The constraints are created by adding artificial spring forces to connect carbon atoms to their original equilibrium positions. By applying such constraints, the "cross-linked" portion will have a relatively constant lattice structures, regardless of the phase transition of the linear PE portion. According to the thermal conductivity of PE fibers (Figure 1a) and those calculated for PEX fibers (see Section S2 in SI), we design the optimal lengths of the linear PE and PEX portions according to Eq. (5) for different fiber diameters (see Section S3 in SI). The predicted thermal rectification factors are listed in Table 1.

*Table 1. Fiber design and prediction of optimized thermal rectification factor*

| Fiber Diameter (nm) | 2.83 | 3.74 | 5.29 | 7.48 |
|---|---|---|---|---|
| Phase Transition Temperature (K) | 317 | 340 | 354 | 366 |
| Predicted rectification factor | 0.76 | 0.96 | 0.96 | 0.86 |

**Figure 3**a illustrates the simulation setup for a PE-PEX junction. Due to the periodic boundary condition, the ends of the linear PE portion are connected to those of the PEX portion, resulting in two interfaces. Two thermostats are added to the middle of the PEX portion and the middle of the linear PE portion, respectively. This allows us to simulate two junctions in one simulation. Only the purple-boxed junction (Figure 3a) is discussed in the following text for convenience, but all reported results are the average value from these two junctions.

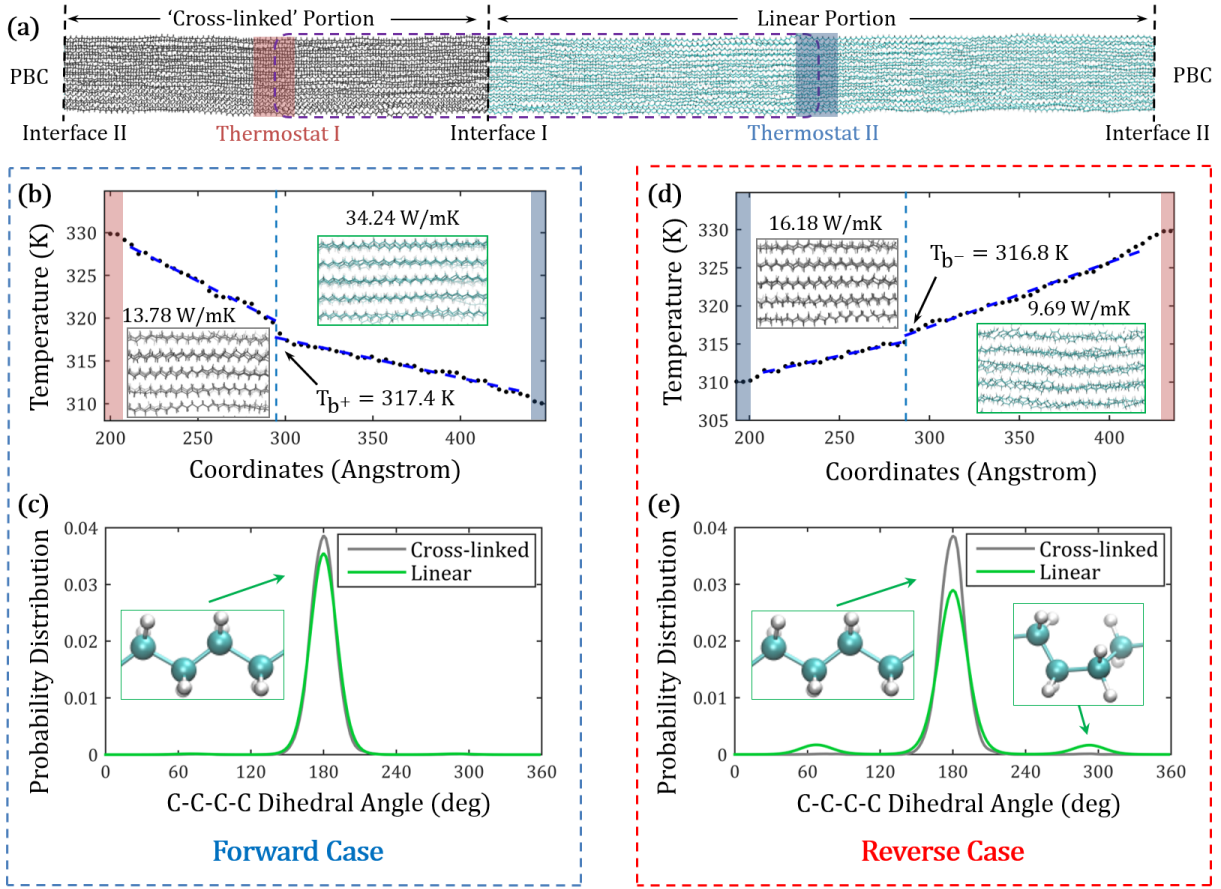

**Figure 3.** a) MD simulation setup of a PE-PEX bi-material junction thermal diode. In the case of forward temperature bias: b) temperature profile with PEX and PE structure snapshots as insets, and (c) dihedral angle distribution of backbone confirming all-*trans* conformation and ordered structures in both PEX and linear PE portions. In the case of reverse temperature bias: d) temperature profile with PEX and PE structure snapshots as insets, and e) dihedral angle distribution of backbone conforming little change of structure in PEX and disordered structures in linear PE portion after the temperature bias is flipped.

In the case of forward temperature bias, thermostat I is set to be the heat source (332 K), and thermostat II is set to be the heat sink (312 K). The heat source and sink temperatures are designed according to the 317 K phase transition temperature of the 2.83 nm PE fiber,[31] ensuring that the phase transition will happen when the temperature bias is reversed. After the heat source and sink are applied, the polymer junction structure with applied heat flux is further relaxed for 1.5 ns using a NPH (isoenthalpic–isobaric) ensemble, which allows the simulation domain size in the chain length direction to relax accordingly at 1 atm pressure. In this forward bias case, the linear PE portion has temperatures lower than the phase transition temperature, and the chain segments are well ordered (right inset, Figure 3b). This is further characterized by analyzing the distribution of the dihedral angles of the carbon backbone (see Section S4 of SI). A sharp mono peak centered at 180° indicates an all-*trans* conformation (Figure 3c). The thermal conductivity of the linear portion is found to be 34.24 W/mK, and that of the PEX portion is 13.78 W/mK. A very small temperature difference is observed at the interface (Figure 3b), and the interfacial thermal conductance is found to be 843 MW/m$^2$K. Considering the lengths and thermal conductivity of each portion, the effective thermal conductivity of the junction is calculated to be 19.70 W/mK.

We then flip the temperature bias, and an equilibration run in NPH ensemble for 1 ns is used to obtain the stable structures for this reverse case. The thermal conductivity of the linear PE portion decreases to 9.69 W/mK. The drop of the thermal conductivity comes from a more disordered structure due to phase change (right inset, Figure 3d), and the dihedral angle distribution is calculated to show that *gauche* conformations have emerged (Figure 3e). Since there is no significant structure change in the PEX portion after the temperature bias is reversed (Figure 3c and 3e), the thermal conductivity of this portion changes much less (16.18 W/mK). In this case, the interfacial thermal resistant is found negligible (Figure 3d). The effective thermal conductivity is calculated to be 11.33 W/mK. The rectification factor from these two cases is 0.74, which agrees very well with the theoretically predicted value of 0.76 (Table 1).

The ratio of the forward to the reverse thermal conductivity can also be calculated simply as the ratio of the heat fluxes in the two cases, since the amplitude of the temperature bias is kept the same. This leads to a rectification factor of 0.78 – agrees well with the value calculated from the effective thermal conductivity. All the following reported rectification factors are calculated using the heat flux ratios. The values are obtained by averaging data from 6 independent simulation runs with the error bars being the standard deviations.

**2.3B Nanoscale size effect on thermal rectification**

PENF with different diameters can have different phase transition temperatures.[31] This will allow the design of thermal diodes that can operate at different temperatures. We study PENFs with diameters ranging from 2.83 nm to 7.48 nm. The lengths of the linear PE and the PEX portions are determined by Eq. (5) in the $\beta = 0$ limit, based on the thermal conductivity contrast ratios from Ref. [[31]]. All the theoretically calculated values are included in Table 1 with more details in Section S3 of SI. The average temperatures used in simulating PE-PEX junctions with different diameters are set according to the corresponding phase transition temperatures (Table 1). The temperature differences are set to be 20K for all cases. **Figure 4**a reports the thermal rectification for fibers with different diameters, highlighting that high thermal rectification factors can be achieved for a wide temperature range (320-380K). The rectification factors from MD simulations also match very well with the theoretical predictions. It is worth noting that the operating temperature of properly designed thermal diodes should be able to extend to ~400K in the thick PE fiber limit since thick PE fibers have a phase transition temperature of 397 K (Figure 1). The thick fiber limit of pure PE was simulated by applying periodic boundary conditions in the cross-sectional directions in our previous work.[31] However, since the PE and the PEX expand differently when temperature changes, the use of periodic boundary conditions in the transverse directions will artificially change the phase transition behavior of the PE portion, and thus rectification cannot be observed as expected. It is possible to simulate an isolated fiber with very large diameters to approach the thick fiber limit. This, however, is restricted by the computational resource. However, based on the thermal conductivity contrast ratio, we can calculate the rectification factor in the thick fiber limit, and it is plotted in Figure 4a at the corresponding phase transition temperature.

It is always possible to obtain larger rectification factors by increasing the thermal bias in most thermal diodes.[5,13] For example, the asymmetric graphene ribbons can have a rectification factor of 3.5, but a temperature difference of 300K is needed.[21] For small temperature differences (*e.g.*, 20 K), the rectification factor of such a device is below 0.2.[21] As suggested by Dames,[13] the rectification factor should be linear in the dimensionless temperature difference, $2(T_H-T_L)/(T_H+T_L)$, to the leading order for almost any rectification mechanism. Thus, to better compare the performances of different rectifiers, it is meaningful to define a dimensionless temperature-scaled thermal rectification factor, $\hat{\varepsilon}$ (Eq. 2). Using such a definition, the PE-PEX thermal diodes show much higher $\hat{\varepsilon}$ than all other diodes (Figure 4b). The PE-PEX diodes have $\hat{\varepsilon}$ ranging from 12 to 25. The above-mentioned asymmetric graphene ribbon diode has a $\hat{\varepsilon}$ of ~4, and $\hat{\varepsilon}$ of almost all the other diodes are found to be smaller than 7.[20-23,41-48] It is worth further noting that the temperature bias of 20K used in this study is by no means the minimal value that can be used in practice. As long as the temperature bias and the fiber lengths are designed properly to enable phase transition, the high thermal rectification factor can be maintained. As a result, $\hat{\varepsilon}$ of the proposed PE-PEX fiber diodes still has room to be improved when temperature bias is further minimized.

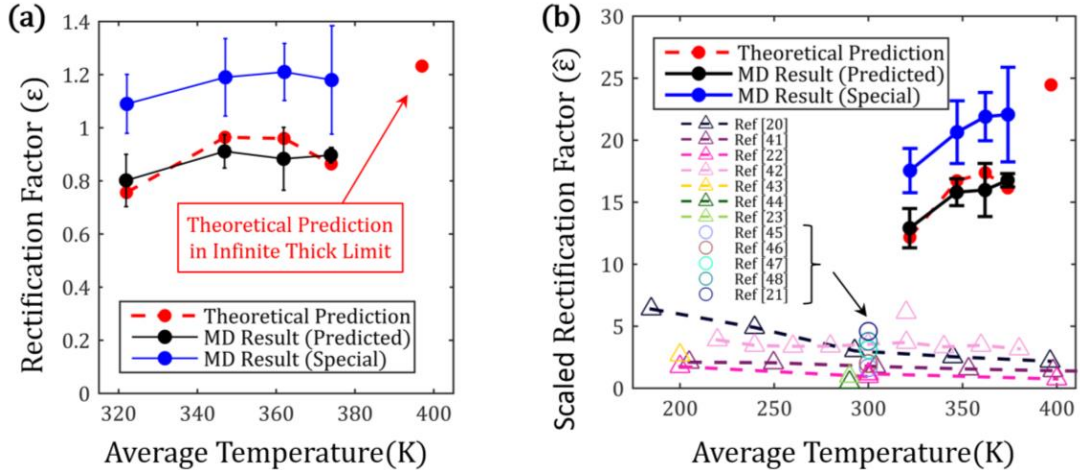

**Figure 4.** a) Thermal rectification factors for PE-PEX fibers with different diameters. b) Dimensionless temperature-scaled thermal rectification factors for PE-PEX diodes and other nanoscale thermal diodes reported in literature.

## 2.3C Higher thermal rectification factor with relaxed constraints

All the above designs require the satisfaction of the condition of $T_{b-} \geq T_{transition} \geq T_{b+}$. However, we found that this requirement can be relaxed to achieve even higher rectification factors. Instead of using lengths predicted according to Eq. (5), if we use lengths tabulated in **Figure 5**a as special cases, the rectification factors are found to be higher than the theoretically predicted upper limit (see blue dots in Figure 4). The reason is that the $\beta \geq 0$ requirement does not necessarily need to be met. This requirement was placed to ensure that the temperature of the whole linear PE portion is completely over or below the phase transition temperature before and after the flip of the temperature bias. In the special cases, it is found $T_{b-} < T_{transition} < T_{b+}$ (Figure 5c and 5d), meaning $\beta < 0$. Here, we analyze why the $\beta \geq 0$ condition can be relaxed. In the forward case, although a small part of the linear PE close to the interface is exposed to temperature higher than the phase transition temperature, phase transition actually does not happen in that part. This is because the configuration of the PEX portion at the interface is not temperature sensitive. It can hold the end of the linear PE portion to maintain its crystalline structure and thus maintain the high thermal conductivity. In the reverse case, when part of the linear PE portion away from the interface undergoes phase transition, the chains begin to rotate, and the rotations can propagate to the part which is still below the phase transition temperature to destroy the entire crystalline structure in the whole linear portion. In real device applications, we can take these factors into consideration, and achieve higher thermal rectification factor by relaxing the $\beta \geq 0$ requirement. As discussed in the theoretical analysis, negative $\beta$ will lead to higher rectification factors, and the predicted value of 1.05 ($\varepsilon_{special}$ in Figure 5b) also agrees quite well with that from the MD simulation (blue point at 322K in Figure 4a).

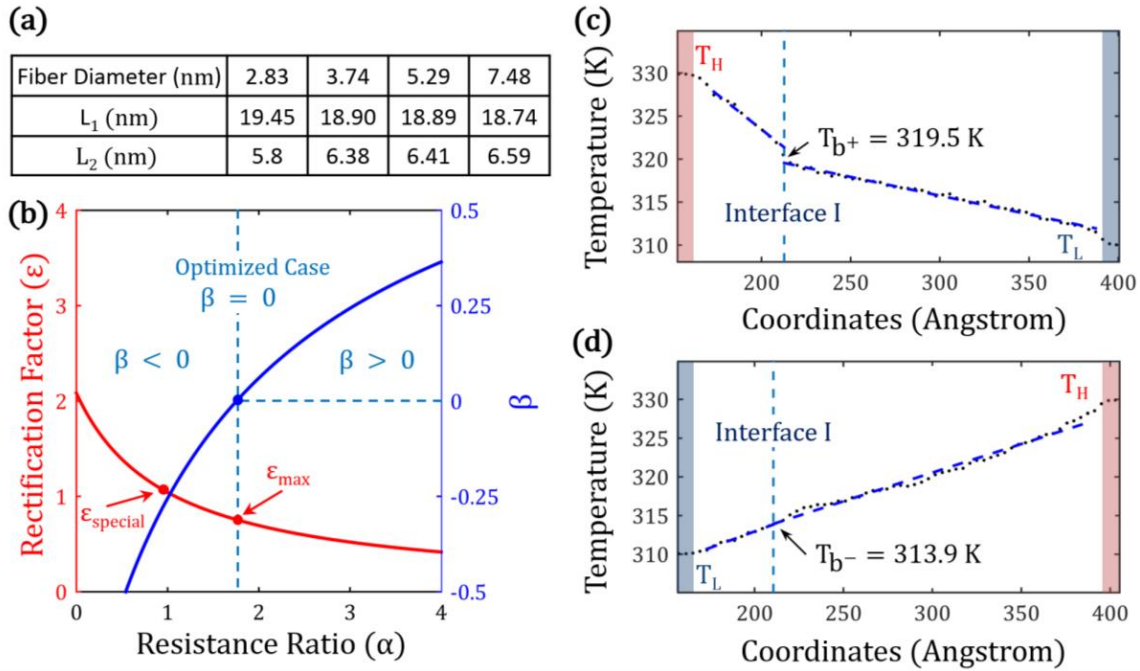

**Figure 5.** a) Special case fiber design. b) Theoretical prediction indicating that $\beta < 0$ can lead to higher rectification factors. c) Forward case and d) reverse case temperature profiles of a 2.83 nm PE-PEX fiber junction.

**2.3D Reproducibility of thermal rectification**

Fully reversible thermal conductivity switching has been observed in PE nanofiber.[31] This should warrant the reversibility of the thermal diode. We here test the reversibility of the thermal diode (flow diagram in **Figure 6**a). By flipping the temperatures at the ends of the forward case, reverse case can be simulated. After NPH relaxation at 1 atm for 2 ns with the reverse temperature setting, NEMD is used to calculate the thermal conductivity ($k^-$). Then, a 15% strain is applied in the chain length direction to help the re-crystallization of the linear PE portion for 0.25 ns, and the temperatures at the ends are flipped again to simulate the forward case. After another 0.25 ns run, the strains are released, and the structures are then relaxed in NPH with 1 atm for 2 ns before the thermal conductivity ($k^+$) is calculated. This completes one test cycle. Seven cycles are tested and a rectification factor ~0.75 is found to be reproducible without degradation observed (Figure 6b). It is worth noting that a strain is needed to achieve the reversibility, because thin PE fibers cannot restore the crystalline phase by themselves.

However, in practice, we can always use a temperature-to-mechanical motion transducer, for example, a bi-layer strip consisting of two layers with different coefficients of thermal expansion, as a holder to hold one end of the fiber to convert the thermal-mechanical device into a pure temperature-controlled device. Moreover, for thick fibers, we have previously shown that the phase transition is reversible without the need of external strain.[31]

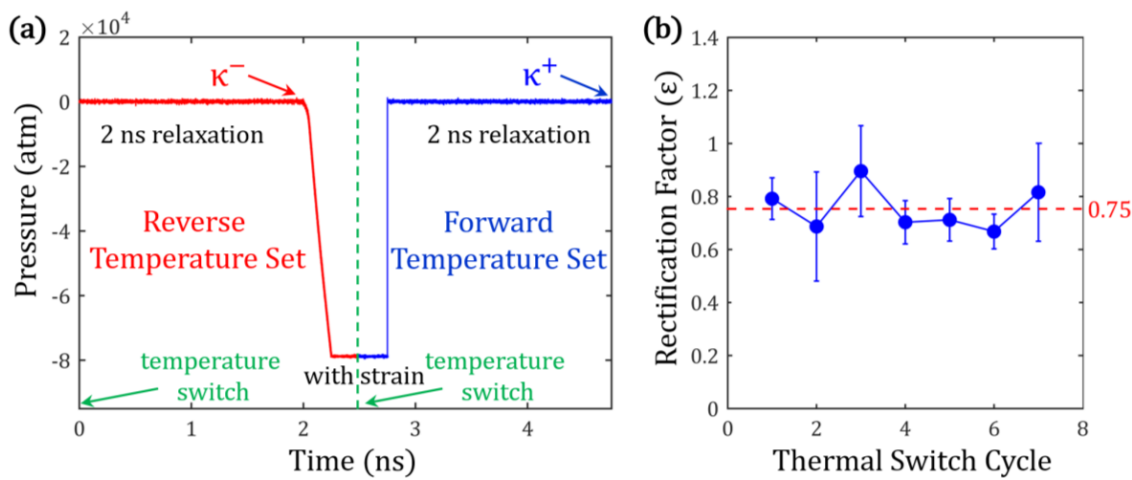

**Figure 6.** a) Flow chart for reversibility test for the 2.83 nm PE-PEX fiber. b) Rectification factors at each cycle.

## 3. Conclusions

Utilizing the phase-transition-induced sharp thermal conductivity switching in PE fibers, thermal didoes are designed by forming a PE-PEX fiber junction. As the temperature bias is flipped, the linear PE portion undergoes phase transition while the PEX portion does not, thus resulting in a significant change of effective thermal conductivity and large thermal rectification. The rectification factors are giant compared to those of the previously reported thermal diodes. MD simulations confirm that the large thermal rectification factor can be achieved with only a 20K temperature difference, which is small enough for real applications. Further MD simulations also demonstrate that the rectification can happen at a wide temperature range and is reversible. Although some challenges are expected to be tackled in the experimental realization of such a device (e.g., fabricating small diameters, proper cross-linking, mechanical durability and so on), the results from this work may open up a new direction for thermal rectification research and provide important guidance (e.g., junction length ratio and fiber size) to achieve practically usable thermal diodes based on inexpensive and easily processable polymer fibers.

## 4. Method

We used the large-scale atomic/molecular massively parallel simulator (LAMMPS)[49] to perform the MD simulations. The condensed-phase optimized molecular potentials for atomistic simulation studies (COMPASS) are used to model the PE structures.[50] A time-step of 0.25 *fs* is chosen due to fast hydrogen vibration. NEMD[51] is used to calculate the thermal conductivity and interfacial thermal conductance of PE structures. As shown in Figure 3a, two Langevin thermostats[52] are applied at each center of the linear PE and PEX portions to impose a temperature gradient across the sample. The temperature gradient ($dT/dx$) is obtained by fitting the linear portion of the temperature profile in steady state (Figure 3b), and heat flux ($J$) is calculated using $J = dQ/dt/S/2$, where dQ/dt is the average of the energy input and output rates in the thermostated regions, and $S$ is the cross-sectional area. The cross-sectional area per chain is defined as the optimized $S$ of the crystal simulation cell divided by the number of chains in the cell, and thus $S$ of a fiber polymer can be calculated by multiplying the number of chains in the fiber. The thermal conductivity ($\kappa$) is calculated by Fourier's law, $\kappa = -J/(dT/dx)$. For the interfacial thermal conductance, temperature drops ($\Delta T$) at the interfaces are determined by the intersection of linear extrapolation of the temperature profile with the interface (Figure 3b). Then, interfacial thermal conductance is calculated by $G = J/\Delta T$.

## Acknowledgements

*The authors acknowledge the financial support from the American Chemistry Society Petroleum Research Fund (54129-DNI10). This research was supported in part by the Notre Dame Center for Research Computing and NSF through TeraGrid resources provided by SDSC Trestles and TACC Stampede under grant number TG-CTS100078.*

# Supporting Information

## S1. Theoretical study of bi-material fiber junction

The proposed thermal diode consists of a portion of linear PE fiber and a portion of PEX fiber (**Figure S1**).

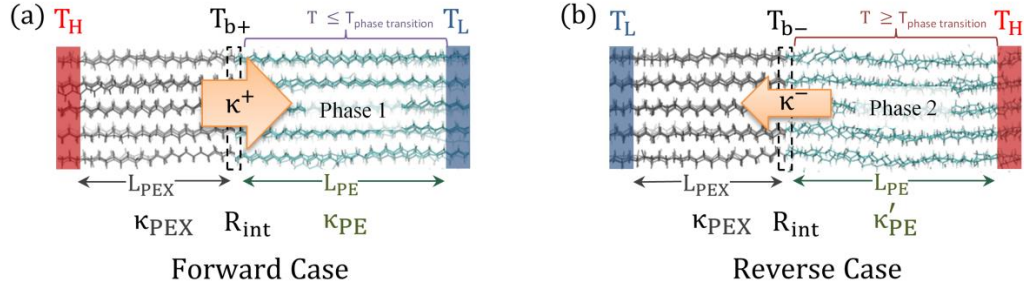

**Figure S1**. Schematics of a thermal diode junction with (a) forward temperature bias and (b) reverse temperature bias. In the case of forward temperature bias, the whole PE portion should be below the phase transition temperature and thus have crystalline structure with higher thermal conductivity. In the case of reverse temperature bias, the whole PE portion should be above the phase transition temperature and thus have disordered structure with lower thermal conductivity.

To simplify the deduction, the fiber lengths ($L_{PE}$ for linear PE, $L_{PEX}$ for PEX), fiber cross sectional area, and the interfacial thermal resistance ($R_{int}$) are considered to remain the same during the switching between forward case and reverse case. The case of forward temperature bias is defined as the larger effective thermal conductivity scheme, in which linear PE is in the higher thermal conductivity ($k_{PE}$) phase. In the reverse case, the linear PE is in the phase of lower thermal conductivity ($k'_{PE}$). Thus, contrast ratio ($\gamma$) of the thermal conductivities $k_{PE}$ to $k'_{PE}$ is always larger than 1. The effective thermal conductivity of the forward case can be written as,

$$k^+ = \frac{L_{PE}+L_{PEX}}{R_{PE}+R_{int}+R_{PEX}} \tag{S1}$$

where $R_{PE}$ is the thermal resistance of the linear PE in the forward case, $R_{int}$ is the interfacial thermal resistance, and $R_{PEX}$ is the thermal resistance of the PEX. Similarly, effective thermal conductivity of the reverse case is:

$$k^- = \frac{L_{PE}+L_{PEX}}{R'_{PE}+R_{int}+R_{PEX}} \tag{S2}$$

where $R'_{PE}$ is the thermal resistance of PE above the phase change temperature. Thus, the ratio ($\sigma$) of the forward and reverse effective thermal conductivity has the following expression:

$$\sigma = \frac{k_+}{k_-} = \frac{R'_{PE}+R_{int}+R_{PEX}}{R_{PE}+R_{int}+R_{PEX}} \tag{S3}$$

With the definitions of $R_{PE} = \frac{L_{PE}}{k_{PE}}$ and $R'_{PE} = \frac{L_{PE}}{k'_{PE}}$, we define $\gamma = \frac{R'_{PE}}{R_{PE}}$. If we let $\alpha = \frac{R_{int}+R_{PEX}}{R_{PE}}$, then the ratio of the forward and reverse effective thermal conductivity ($\sigma$) can be written as a function of only $\gamma$ and $\alpha$:

$$\sigma = \frac{\alpha+\gamma}{\alpha+1} \tag{S4}$$

Since $\alpha$ is positive and $\gamma$ is large than 1, $\sigma$ will increase monotonically with respect to $\gamma$ and decrease monotonically with respect to $\alpha$. In addition, rectification factor ($\varepsilon$) is a function of $\sigma$, and increases monotonically with $\sigma$:

$$\varepsilon = \sigma - 1 \tag{S5}$$

From the above discussion, it can be concluded that larger $\gamma$ and smaller $\alpha$ give larger $\varepsilon$.

However, the possible value of $\gamma$ is limited by the thermal conductivities of the materials in the two phases. $\gamma$ can be as large as ~12 in thick PE crystalline fibers with a temperature difference of 150K.[31] For a temperature difference of 20K, $\gamma$ is ~5 for thick crystalline PE fiber, and ~3 for PENF.[31] Although smaller $\alpha$ is always desired, $\alpha$ cannot be infinitesimal. In the forward case, $T_{b+}$ must be below the phase transition temperature ($T_{phase\ transtion}$), and thus the whole linear PE portion can maintain the high thermal conductivity phase (Figure S1a). In the reverse case, $T_{b-}$ must be above $T_{phase\ transtion}$, and thus the phase transition to disordered structures with low thermal conductivity can happen in the whole linear PE portion (Figure S1b). The temperature constraint ($T_{b-} \geq T_{b+}$) at the interface can also be expressed as a function of $\alpha$ and $\gamma$:

$$\beta = \frac{T_{b-}-T_{b+}}{\Delta T} = \left(\frac{T_{b-}-T_L}{T_H-T_L} - \frac{T_{b-}-T_L}{T_H-T_L}\right) = \left(\frac{R_{int}+R_{PEX}}{R'_{PE}+R_{int}+R_{PEX}} - \frac{R_{PE}}{R_{PE}+R_{int}+R_{PEX}}\right) = \left(\frac{\alpha}{\alpha+\gamma} - \frac{1}{\alpha+1}\right) \geq 0 \tag{S6}$$

In the limit of $\beta = 0$, the smallest allowable $\alpha$ is achieved, and optimized solution for the largest possible $\varepsilon$ can be obtained:

$$\alpha = \gamma^{0.5}, \sigma\ (optimized) = \frac{\gamma^{0.5}+\gamma}{\gamma^{0.5}+1}\ \text{and}\ \varepsilon\ (optimized) = \frac{\gamma-1}{\gamma^{0.5}+1}. \tag{S7}$$

**S2. Thermal conductivity of PEX**

In this section, we show the calculated thermal conductivity of PEX with different lengths and diameters at different temperatures. These data are used for designing the length of each portion in the PE-PEX thermal diode junction for MD simulations. The fiber diameters studied are 2.83 nm, 3.74 nm, 5.29 and 7.48 nm. The total lengths of the fibers are around 50 nm (400 -$CH_2$- segments). For the "cross-linked" PE fiber, constraints are created by adding artificial spring forces to connect carbon atoms to their original equilibrium positions, and the force constant is set to be 599.34 Kcal/mole/Å$^2$. By applying such constraints, the "cross-linked" will have a relatively constant cross-sectional lattice and along-chain order, regardless of the phase transition of linear PE portion (Figure 3 in main text).

For the design of the PE-PEX thermal diode, the thermal conductivities of these PEX fibers are first studied as a function of temperature. For a fiber of 50 nm in length and 2.83 nm in diameter, the thermal conductivity of linear PE shows sharp thermal conductivity drop at ~320K, but the thermal conductivity of PEX remains relatively constant over the whole studied temperature range 300-340K (**Figure S2**a). This temperature-independent thermal conductivity in PEX comes from the reason that the fixed-spring constraints in our "cross-linked" PE stabilize the polymer structures. The suppression of the phase transition explains the relatively constant thermal conductivity. In real PEX, chemical bonds will connect different PE single chains, stabilizing the polymer structures, and thus high thermal stability without phase transition is possible in PEX of high cross-link density.

Then, we calculate the thermal conductivity of PEX fiber with a diameter of 2.83 nm as a function of the fiber length (Figure S2b). The thermal conductivity of PEX fiber is found to increase as fiber becomes longer. We choose 11.5 W/mK as a rough estimation for the PE-PEX thermal diode design. In addition, the MD simulation results of PEX thermal conductivities in the designed PE-PEX junction are 13.78 W/mK and 16.18 W/mK, which are close to our 11.5 W/mK estimation (Figure 3 in the main text). Figure S2c also shows that the thermal conductivity of PEX is not a strong function of the fiber diameter, and thus the same value (11.5 W/mK) is used for thermal diode design of all different fiber diameters.

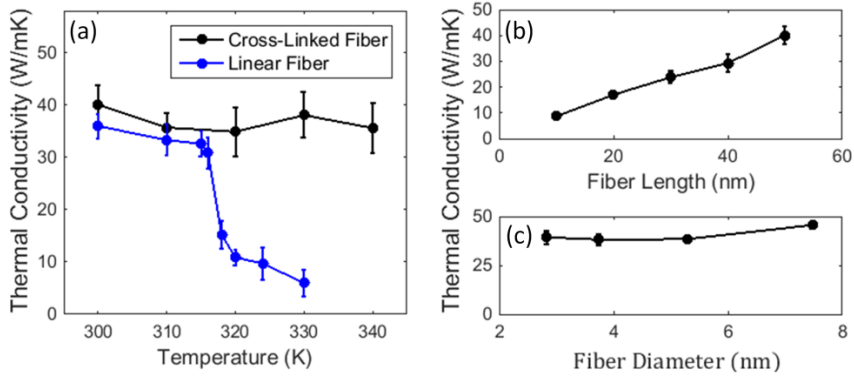

**Figure S2**. (a) Thermal conductivity of PEX and linear PE fibers (length = 50nm, diameter = 2.83nm) from 300K to340K. (b)Thermal conductivity of PEX fiber of different lengths (300K, diameter = 2.83nm). (c) Thermal conductivity of PEX fiber of different diameters (300K, length = 50nm).

### S3. Design of PE-PEX thermal diode junction and theoretical thermal rectification factor

Our MD simulations of thermal diodes need to start from certain thermal diode design. To find out the optimized length ratio of PEX to linear PE ($L_{PEX}/L_{PE}$), we base our thermal diode design on the theoretical analysis in Section S1, using the known thermal conductivities ($\kappa_{PE}$ and $\kappa'_{PE}$) of linear PE fibers of different diameters in different phases from ref. [31] and the estimated thermal conductivities ($\kappa_{PEX}$) of PEX portion from the above section. We estimate the optimal lengths of the linear PE and PEX portions according to Eq. (5) in the main text. In the $\beta = 0$ limit, $\alpha_{\min}$ can be solved as a function of only $\gamma$ according to Eq. (4) in the main text, and $\gamma$ is calculated based on the thermal conductivity data of PE fibers before and after phase transition (Table S1). The thermal conductivity of the PEX fiber with a diameter of 2.83 nm is calculated using NEMD separately, and the value is listed in Table S1. The same value is used for all diameters since we found that diameter does not influence the thermal conductivity significantly (Figure S2c). According to Eq. (5) in the main text, the proper PEX length is determined to be 9.8 nm, and the linear PE is 15.4 nm for a nanofiber of 25.2 nm in total length (Table S1). Please note that the total simulation domain size is ~50 nm which contains two junctions connected end-to-end due to periodic boundary condition in the along-chain direction (Figure 3a in main text). Each junction thus has a length of ~25 nm.

*Table S1. Fiber design and prediction for optimized thermal rectification factor*

| Fiber Diameter (nm) | 2.83 | 3.74 | 5.29 | 7.48 |
|---|---|---|---|---|
| Chain Number | 40 | 70 | 140 | 280 |
| Phase Transition Temperature (K) | 317 | 340 | 354 | 366 |
| $\kappa_{PE}$ | 31.70 | 31.63 | 30.53 | 27.29 |
| $\kappa'_{PE}$ | 10.28 | 8.20 | 7.95 | 7.85 |
| $\kappa_{PEX}$ | 11.5 | 11.5 | 11.5 | 11.5 |
| $L_{PEX}/L_{PE}$ | 0.64 | 0.71 | 0.74 | 0.79 |
| $L_{PE}$ (nm) | 15.4 | 14.8 | 14.6 | 14.0 |
| $L_{PEX}$ (nm) | 9.8 | 10.5 | 10.7 | 11.1 |
| Predicted | 0.76 | 0.96 | 0.96 | 0.86 |

**S4. Structure characterization of PE**

The thermal conductivity of PE is highly morphology-dependent, as shown in our previous papers.[30,31] Generally, phonon transport in solid materials with perfect lattice is of high efficiency due to less disorder phonon scattering. For the thermal transport along polymer chains, the thermal conductivity is significantly influenced by the along chain order.[30, 31] Since the dihedral angle strength is weak, the segments in the chain can rotate and, leading to disorders along the polymer chain. These disorders will serve as phonon scattering sites and thus lead to dramatic drop of the thermal conductivity (**Figure S3**). At the molecular level, all-*trans* conformation chains are found in crystalline PE. *Gauche* conformations will occur when segments rotate (Figure S3).

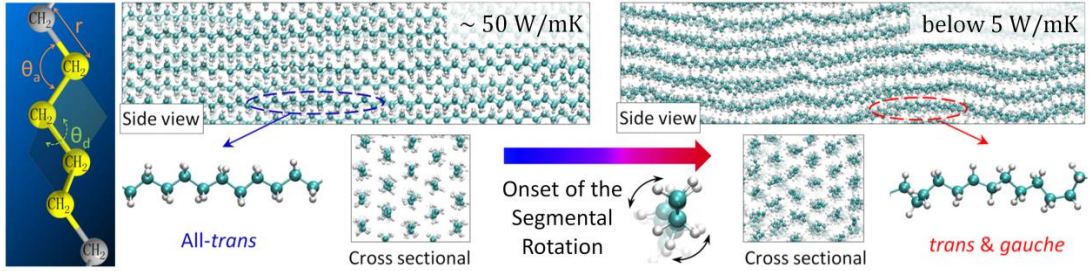

**Figure S3**. The morphology of polymer is largely dominated by segment rotations, since dihedral angle strength is the weakest interaction among all kinds of intramolecular interactions. The onset of the segment rotation changes all-trans conformation crystalline phase to disordered phase with random segmental rotations, and dramatically lowers the thermal conductivity in the chain length direction.

As a result, the most direct characterization of the along-chain polymer morphology is to calculate the dihedral angle distribution of backbone, since *trans* and *gauche* conformations can be directly distinguished as different peaks in the distribution (**Figure S4**).

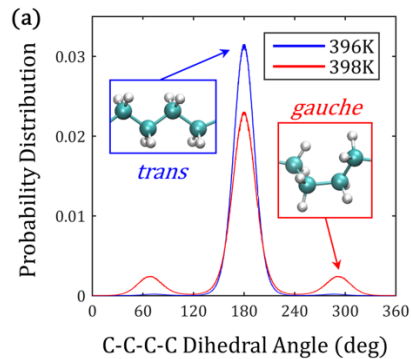

**Figure S4**. Structure characterization of thick PE fiber: dihedral angle distribution of backbone directly presents the structure change from all-trans conformations to a combination of trans and gauche conformations.